\documentclass[showpacs,pre]{revtex4}
\usepackage{graphicx}
\begin{document}
\title{Delocalization transition of a small number of particles in a box with periodic boundary conditions}
\author{Hidetsugu Sakaguchi}
\affiliation{Department of Applied Science for Electronics and Materials,
Interdisciplinary Graduate School of Engineering Sciences, Kyushu
University, Kasuga, Fukuoka 816-8580}
\begin{abstract}
We perform molecular dynamics simulation of a small number of particles in a box with periodic boundary conditions from a view point of chaotic dynamical systems. There is a transition at a critical energy $E_c$  that each particle is confined in each unit cell for $E<E_c$, and the chaotic diffusion occurs for $E>E_c$. We find an anomalous behavior of the jump frequency above the critical energy in a two-particle system, which is related with the infinitely alternating stability change of the straight motion passing through a saddle point.  
 We find simultaneous jump motions just above the critical energy in a four-particle system and sixteen-particle system, which is also related with the motion passing through the saddle point.
\end{abstract}
\maketitle
\section{Introduction}
Liquid-solid phase transitions have been long studied by the molecular dynamics (MD) simulation, since Alder and Wainwright found the liquid-solid phase transition in a system of hard spheres~\cite{rf:1}.  Various thermodynamic properties have been studied with numerical simulations of a large number of particles~\cite{rf:2}. Alder and Wainwright also found the long-time tail of the velocity auto-correlation, which leads the divergence of the transport coefficient such as the diffusion constant and the viscosity in one and two dimensions~\cite{rf:3}. 
Recently, it is  considered that dynamical heterogeneities play an important role in the glass transition and the jamming transition of granular material~\cite{rf:4}. Successive jump motions and intermittent swirling motions are observed in MD simulations of supercooled liquids~\cite{rf:5,rf:6,rf:7}.
A deterministic chaotic dynamics is assumed in the MD simulation, and therefore, the liquid-solid phase transition might be interpreted as a kind of dynamical transition in chaotic systems. 
The liquid -solid phase transition was studied using the Lyapunov spectrum which characterizes the chaotic dynamics by Posch et al.  The most-positive Lyapunov exponent exhibits a maximum at the phase transition~\cite{rf:8,rf:9}.  On the other hand,  Awazu and Munakata-Hu studied a system of only two hard disks in a rectangular box~\cite{rf:10,rf:11} They showed that there is a dynamical transition similar to a liquid-solid phase transition in the system of a small number of particles.
.

In this paper, we study numerically a similar dynamical transition in Hamiltonian systems of a small number of particles in a box with periodic boundary conditions. That is, each particle is confined in each unit cell for $E<E_c$, and the chaotic diffusion occurs for $E>E_c$. We call it a delocalization transition in this paper.   The delocalization transition in a small number of particles has a similarity to the liquid-solid phase transition in an infinite-size system, in that each particle is confined in each unit cell in the solid phase and the particles move around beyond the unit cells in the liquid phase. However, the relation between the delocalization transition and the statistical-mechanical liquid-solid phase transition is not clear now.  The purpose of this paper is to show some peculiar behaviors found near the delocalization transition.
 
In \S 2, we study a two-particle system, and discuss an anomalous behavior of the chaotic diffusion above the critical energy. In \S 3, we study a four-particle system and a sixteen-particle system, and discuss simultaneous jump motions just above the critical energy. We will relate these behaviors with the motion passing through a saddle point. 

We consider Hamiltonian systems interacting with repulsive forces. The repulsive force between two particles  at ${\bf r}_1$ and ${\bf r}_2$ is assumed to be $\bf{F}(r)=-(\partial U/\partial r)\cdot {\bf r}/|{\bf r}|$ where ${\bf r}={\bf r}_2-{\bf r}_1$, using the Lennard-Jones potential $U(r)$:
\begin{equation}
U(r)=e\left \{\left (\frac{\sigma}{r}\right )^{12}-\left(\frac{\sigma}{r}\right)^6\right\},
\end{equation}
for $r<r_0$ with $\sigma=r_0/(2^{1/6})$. The attractive part of the Lennard-Jones potential is not used, that is, $F(r)$ is assumed to be 0 for $r>r_0$.
The repulsive force is therefore a short-range force, and $U(r)$ is a monotonically decreasing function of $r$. For our numerical simulation, $r_0$ is set to be 1, $e=0.01$ and the mass $m$ is assumed to be 1 for all particles.  The fourth order symplectic method of timestep $\Delta t=0.0001$ was used for the numerical simulation.
\begin{figure}[tbp]
\includegraphics[height=4.5cm]{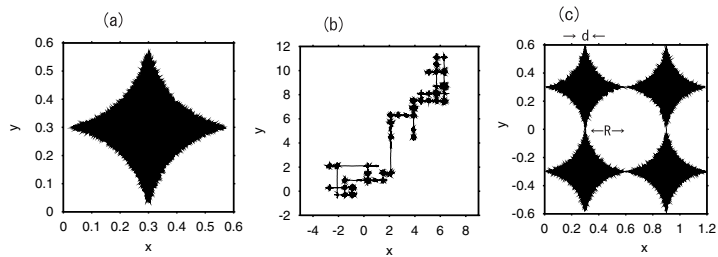}
\caption{Trajectories of the first particle at (a) $E=1.992$ and (b) $E=2.228$ for $L=1.2$. (c) Trajectory of the first particle at  $E=2.104$. The radius of a circle surrounded by the four shaded regions is denoted by $R$ and the gate width is denoted by $d$.}
\label{fig1}
\end{figure}
\begin{figure}[tbp]
\includegraphics[height=4.cm]{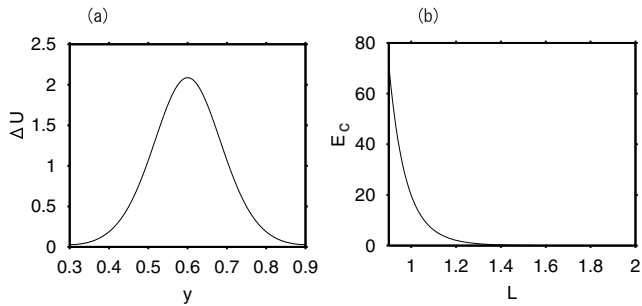}
\caption{(a) Potential energy along a trajectory $(L/4,y)$ of the first particle for $L=1.2$. (b) Critical energy of the delocalization transition as a function of $L$.}
\label{fig2}
\end{figure}
\section{Anomalous behavior of chaotic diffusion in a two-particle system} 
As one of the simplest system, we consider a two-particle system in a square box.  The two particles are further assumed to be located at point-symmetric positions with respect to the center $(L/2,L/2)$ of the square box. 
 That is, the position ${\bf r}_2$ of the second particle  is represented as ${\bf r}_2=(x_2,y_2)=(L-x_1,L-y_1)$ by the position ${\bf r}_1=(x_1,y_1)$ of the first particle.  The periodic boundary conditions are further assumed. This system is equivalent to a system that the pair of particles are arranged periodically in space with period $L$. The equations of motion for the first particle are given by
\begin{eqnarray}
\frac{d^2x_1}{dt^2}&=&-\sum_{i,j}e\left \{6\left (\frac{\sigma}{r_{ij}}\right )^6-12 \left (\frac{\sigma}{r_{ij}}\right )^{12}\right \}\frac{x_1-x^{\prime}_i}{r_{ij}^2},\nonumber\\ 
\frac{d^2y_1}{dt^2}&=&-\sum_{i,j}e\left \{6\left (\frac{\sigma}{r_{ij}}\right )^6-12 \left (\frac{\sigma}{r_{ij}}\right )^{12}\right \}\frac{y_1-y^{\prime}_j}{r_{ij}^2},
\end{eqnarray}
where $(x^{\prime}_i,y^{\prime}_j)=(x_2+i\cdot L,y_2+j\cdot L)$ and $r_{ij}=\sqrt{(x_1-x^{\prime}_i)^2+(y_1-y^{\prime}_j)^2}$.
The summation is taken only for the particles at $(x^{\prime}_i,y^{\prime}_j)$ satisfying $r_{ij}\le r_0$. 
This system is a Hamiltonian system with four degrees of freedom of motion and conserves the total energy $E$. In this paper, the potential energy is counted as $\Delta U=U-U(r_0)$ from the minimum value $U(r_0)$. Our system is one of the minimum system that can exhibit the delocalization transition. Figures 1(a) and (b) display the trajectories of $(x_1,y_1)$ for (a) $E=1.992$ and (b) $E=2.228$ at $L=1.2$.  The initial conditions are slightly different, i.e., (a) $x_1(0)=y_1(0)=L/4,dx_1(0)/dt=0.05,dy_1(0)/dt=1.4$ and (b) $x_1(0)=y_1(0)=L/4,dx_1(0)/dt=0.05,dy_1(0)/dt=1.5$. Note that the ranges of the $x$ and $y$ coordinates are very different. 
For $E=1.992$, the particle exhibits chaotic motion but it is confined in a square box of $[0,L/2]\times [0,L/2]$. However, for $E=2.228$, the particle exhibits chaotic diffusion. The $(x,y)$ space is divided into cells with unit length $L/2$. The particle is confined inside of one cell when $E$ is smaller than a critical value $E_c$, however, the particle moves around in a chaotic manner from one cell to another cell for $E>E_c$.  This transition is called a delocalization transition.

\begin{figure}[tbp]
\includegraphics[height=4.cm]{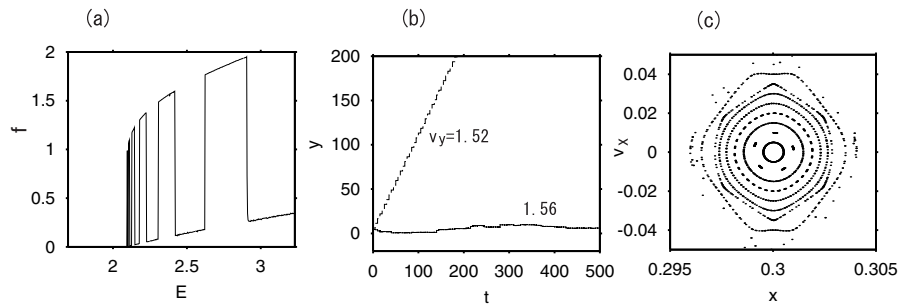}
\caption{(a) Frequency $f$ that the particle passes through the periodic boundaries as a function of $E$ for $L=1.2$. (b) Time evolution of $y_1(t)$ for two initial velocities $dy_1(0)/dt=1.52$ (dashed line) and 1.56 (solid curve). (c) Poincare plot in the $x_1$-$v_x$ space at the section $y_1=L/4$. The initial conditions  are $v_x(0)=0.005\times n$ ($n=1,2,3\cdots$ 10.), $v_y(0)=1.52,x(0)=L/4$ and $y(0)=L/4$.}
\label{fig3}
\end{figure}
\begin{figure}[tbp]
\includegraphics[height=4.cm]{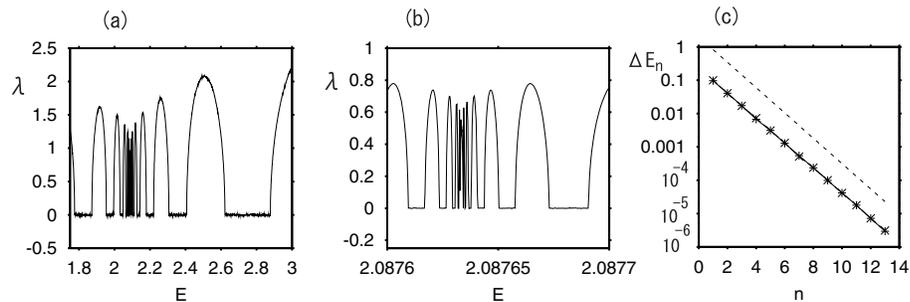}
\caption{(a) Linear growth rate $\lambda$ of $|\delta x_1|$ as a function of $E$ for $L=1.2$ in the range between $1.75<E<3$. (b)  Linear growth rate $\lambda$  in the range between $2.0876<E<2.0877$.(c) Width $\Delta E_n$ of the $n$th parameter range of $E$ where $\lambda$ is nearly 0. The dashed line denotes a line of $\Delta E_n\propto 1/(2.4)^n$.}
\label{fig4}
\end{figure}
\begin{figure}[tbp]
\includegraphics[height=4.cm]{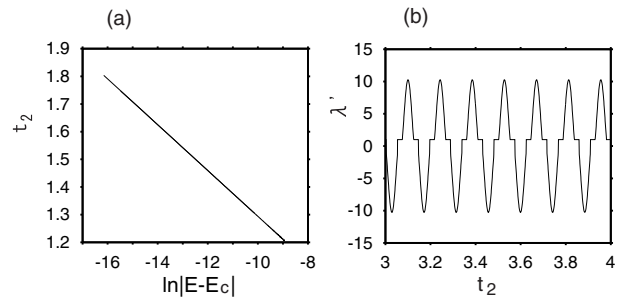}
\caption{(a) Period $t_2$ of $y_1(t)$ in Eq.~(3) as a function of $\ln|E-E_c|$. (b) Stability exponent $\lambda^{\prime}$ as a function of the period $t_2$ for a simple model Eq.~(5).}
\label{fig5}
\end{figure}
 The point $(L/4,L/4)$ is a point where the potential energy takes the minimum. On the other hand, the point $(L/4,L/2)$ is a saddle point of the potential energy, because the potential $U(L/4,y)$ takes a maximum value at $y=L/2$ along a line $x=L/4$ but $U(x,L/2)$ takes a minimum value at $x=L/4$ along a line $y=L/2$. Figure 2(a) displays the potential energy $\Delta U(y)=U({\bf r}_1)-U(r_0)$ along a line $(L/4,y)$. The minimum energy to get over the saddle point is therefore calculated as $E_c=2\{U(L/2)-U(r_0)\}$, because particles located at $(3L/4,L/2)$ and $(-L/4,L/2)$ interact with the first particle at the saddle point $(L/4,L/2)$.  
It is a critical energy for the delocalization transition. 
Figure 2(b) displays the critical energy $E_c$ as a function of the system size $L$.  As $L$ is decreased, the critical energy $E_c$ increases rapidly, because the two particles are confined in a smaller box.   

The shaded region in Fig.~1(a) is surrounded by four arcs whose radius $R$ satisfies $2\{U(2R)-U(r_0)\}=E$. As $E$ is increased, $R$ decreases, because $U$ is a monotonically decreasing function.  For $E<E_c$, $R$ is larger than $L/4$. Then, the shaded region surrounded by the four arcs is confined in the cell of size $L/2$, and the chaotic motion is also confined inside of the unit cell region. At $E=E_c$, $R$ is equal to $L/4$ and the four arcs touch the saddle points at $y=L/2$ or $x=L/2$ and the chaotic motion invades into the neighboring cells, and the chaotic diffusion appears. For $E>E_c$, the shaded region surrounded by the four arcs is connected with the shaded region in the neighboring cell as shown in Fig.~1(c) at $E=2.104$. The arcs of the neighboring four cells form a circle of radius $R$. The gate width $d$ formed by the overlap of the neighboring shaded regions at the section $y=L/2$ is equal to $d=L/2-2R$, which is proportional to $E-E_c$ near the critical energy $E_c$.  The particle moves to the neighboring cells through these narrow gates with gate width $d$ for $E>E_c$.      

We have numerically calculated the frequency $f$ that the first particle passes through the periodic boundaries $x_1=n_xL$ or $y_1=n_yL$ ($n_x$ and $n_y$ are integers).  The initial condition is  $x_1(0)=y_1(0)=L/4,dx_1(0)/dt=0.001$ and the initial velocity $dy_1(0)/dt$ is changed as a parameter to change the total energy $E$. 
Figure 3(a) displays the frequency $f$ as a function of the energy $E$ for $L=1.2$. 
The frequency $f$ is 0 for $E<E_c\sim 2.09$, because the particle is confined inside of the cell as shown in Fig.~1(a). For $E>E_c$, $f(E)$ takes a nonzero value, however, the behavior of $f(E)$ is rather anomalous. Roughly speaking, there are two branches for $f(E)$, and which branch is selected depends on the energy.  In the lower branch of $f(E)$, $f(E)$ increases  as $f\propto |E-E_c|$ near the critical energy, which corresponds to the chaotic diffusion as shown in Fig.~1(b).  It is because the frequency $f$ is proportional to the gate width $d$ formed by the overlap of the neighboring shaded regions, and therefore $f$ is proportional to $E-E_c$ near the critical energy $E_c$.

In the energy intervals where $f(E)$ takes the upper branches in Fig.~3(a), the particle exhibits an almost straight motion in the $y$ direction. This mode of motion is called the accelerator mode. This type accelerator modes were studied in various systems such as the standard map or the chaotic diffusion in an oscillatory B\'enard convection~\cite{rf:12,rf:13}. 
The anomalous chaotic diffusion in a dissipative system was also studied in an oscillatory B\'enard convection~\cite{rf:14}. Figure 3(b) displays time evolutions of $y_1(t)$ for two initial velocities $dy_1(0)/dt=1.52$ and 1.56. The other initial values are the same as before, i.e.,  $x_1(0)=y_1(0)=L/4,dx_1(0)/dt=0.001$.  For $v_y(0)=1.52$, $y_1(t)$ increases in proportion to $t$, which implies the straight motion in the $y$-direction. For $v_y(0)=1.56$, $y_1(t)$ exhibits chaotic motion around $y_1=0$. Figure 3(c) displays the Poincare plot in the $x$-$v_x$ space at the section $y_1=L/4$ for the initial conditions $v_x(0)=0.005\times n$ ($n$ is an integer), $v_y(0)=1.52,x(0)=L/4$ and $y(0)=L/4$. A straight motion of $x_1(t)=L/4=$const. corresponds to a central fixed point in this Poincare map. The KAM tori around this fixed point represent stable accelerator modes. 
A large sea of chaos exists around the small region of stable KAM tori.  
If the initial conditions are in the KAM region, the particle exhibits a straight motion, and if the initial conditions are outside of the KAM region, the particle exhibits chaotic diffusion. If the initial conditions are located in the intermediate region, the straight motion appears intermittently and the diffusion constant becomes large as shown in [12],[13] and [14].

The completely straight motion along the line $x=L/4$ is a special solution. 
The straight motion is stable for some parameter ranges where the KAM tori appear as shown in Fig.~3(c).
The stability of the straight motion can be studied from the time evolution of the small deviation $\delta x(t)$ from the straight motion along the line $x=L/4$. The $y$-coordinate $y_1(t)$ of the straight motion and the deviation $\delta x(t)$ obey
\begin{eqnarray}
\frac{d^2y_1}{dt^2}&=&-\sum_{i,j}e\left \{6\left (\frac{\sigma}{r_{ij}}\right )^6-12 \left (\frac{\sigma}{r_{ij}}\right )^{12}\right \}\frac{y_1-y^{\prime}_j}{r_{ij}^2},\\
\frac{d^2\delta x_1}{dt^2}&=&-\sum_{i,j}e\left \{6\left (\frac{\sigma}{r_{ij}}\right )^6-12 \left (\frac{\sigma}{r_{ij}}\right )^{12}\right \}\frac{2\delta x_1}{r_{ij}^2}\nonumber\\
& &+\sum_{i,j} e\left \{48\left (\frac{\sigma}{r_{ij}}\right )^6-168 \left (\frac{\sigma}{r_{ij}}\right )^{12}\right \}\frac{2(L/4-x_j)^2\delta x_1}{r_{ij}^4}. 
\end{eqnarray}
The motion of $y_1(t)$ is periodic in time and the period increases to infinity at the critical energy $E_c$.
The transverse instability is measured by the linear growth rate $\lambda$ of $|\delta x_1(t)|$. Figure 4(a) displays $\lambda$ as a function of $E$ in the range $1.75<E<3$. The unstable parameter regions and the stable regions appear alternatively. The unstable parameter regions for $E>E_c$  correspond to the parameter regions where the lower branch of $f(E)$ appears in Fig.~3(a) and the stable parameter regions for $E>E_c$ correspond to the parameter regions of the upper branch of $f(E)$ in Fig.~3(a). The width $\Delta E$ of each parameter region decreases as $E$ approaches $E_c$. Figure 4(b) displays the magnification of Fig.~4(a) in the range $2.0876<E<2.0877$. The alternation of the stability continues infinitely as $E\rightarrow E_c$ both for $E<E_c$ and $E>E_c$. Figure 4(c) displays the width $\Delta E_n$ of the $n$th stable parameter ranges as a function of $n$ in a semi-logarithmic scale. The width $\Delta E_n$  for $E<E_c$ is marked by $+$ and the one for $E>E_c$ is marked by $\times$, although they are almost overlapped in Fig.~4(c).  Figure 4(c) implies that the width $\Delta E_n$ decreases exponentially as $\Delta E\propto (1/2.4)^n$ as $E$ approaches $E_c$. 

To the best of our knowledge, this type of singular behavior was not reported before.  This behavior can be qualitatively understood with a simpler model equation. The equation for $\delta x_1$ is roughly approximated as 
\begin{eqnarray}
\frac{d^2\delta x_1}{dt^2}+\omega_1^2\delta x_1&=&0,\;\;\;{\rm for }\;\; 0<t<t_1,\nonumber\\   
\frac{d^2\delta x_1}{dt^2}+\omega_2^2\delta x_1&=&0,\;\;\;{\rm for }\;\; t_1<t<t_2,
\end{eqnarray}
where $t_2$ is the period of $y_1$ by Eq.~(3), $t_2-t_1$ is the time during which the particle stays near the saddle point $y_1=L/2$, and $t_1$ is the time during which the particle is far away from the saddle point. 
That is, the complex function of the left-hand side of the equation for $\delta x_1$ in Eq.~(4) is approximated at a piecewise linear equation (5), which characterizes respectively the dynamical behavior near and far from the saddle point. 
As $E$ is close to $E_c$, the period $t_2$ increases in a logarithmic manner as $t_2\sim -\alpha\ln|E-E_c|$. The approximation by the second equation in Eq.~(5) would be better near the critical energy $E_c$, because $t_2-t_1$ increases infinitely and the particle stays near the saddle point for a very long time. 
Figure 5(a) displays $t_2$ as a function of $\ln |E-E_c|$ by the direct numerical simulation of Eq.~(3). The parameter $\alpha$ is evaluated as $\alpha\sim 0.0827$. The parameter $\alpha$ is related to the behavior of the potential $\Delta U$ near the saddle point shown in Fig.~2(a), i.e., $\Delta U\sim E_c-1/(2\alpha^2)(y-L/2)^2$.  
The stability of the linear equation (5) per one period $t_2$ can be 
calculated from eigenvalues of the matrix $C=AB$ where
\[
A=\left (
\begin{array}{cc}
\cos\omega_2t_2 & \sin\omega_2 t_2\\
-(\omega_2/\omega_1)\sin\omega_2t_2 & (\omega_2/\omega_1)\cos\omega_2t_2
\end{array}\right ),\]
\[
B=\left (
\begin{array}{cc}
\cos\omega_1t_1\cos\omega_2t_1+(\omega_1/\omega_2)\sin\omega_1t_1\sin\omega_2t_1 &
\sin\omega_1t_1\cos\omega_2t_1-(\omega_1/\omega_2)\cos\omega_1t_1\sin\omega_2t_1\\
\cos\omega_1t_1\sin\omega_2t_1-(\omega_1/\omega_2)\sin\omega_1t_1\cos\omega_2t_1 &
\sin\omega_1t_1\sin\omega_2t_1+(\omega_1/\omega_2)\cos\omega_1t_1\cos\omega_2t_1\end{array}
\right ).
\]
The largest eigenvalue $\lambda^{\prime}$ of $C$ expresses the stability of the linear equation (5) per one period $t_2$.  Figure 5(b) displays the largest eigenvalue $\lambda^{\prime}$ if it is a real number, or the modulus of $\lambda^{\prime}$ if the eigenvalues are complex number, as a function of $t_2$ for $\omega_1^2\sim 100$ and $\omega_2^2=1940.7$. The growth of the deviation $|\delta x_1(t)|$ in a large time interval $t$ is evaluated as $|\delta x_1(t)|\sim |\lambda^{\prime}|^{t/t_2}\sim \exp(\lambda t)$, because the particle passes through the saddle point $(t/t_2)$ times in the time interval $t$. Therefore, the exponent $\lambda$ for the transverse instability is evaluated as $\lambda\sim \ln|\lambda^{\prime}|/t_2$. 
Here, the value $\omega_2^2=1940.7$ is evaluated from the right-hand side of Eq.~(4) at the saddle point $(L/4,L/2)$ near $E=E_c$, and $\omega_1^2$ is roughly estimated from the right-hand side of Eq.~(4) near the fixed point $(L/4,L/4)$. If $|\lambda^{\prime}|>1$, $\delta x_1$ increases to infinity and the straight motion is unstable. Figure 5(b) shows that the stability alternates periodically as a function of the period $t_2$. The period $T$ of the alternation of the stability is equal to $\pi/\omega_2\sim 0.0713$, because of the periodicity of the matrix $A$. (The sign of $\lambda^{\prime}$ has no meaning.)  The period $t_2$ of the straight motion is evaluated as $t_2=-\alpha\ln|E_n-E_c|$ at the $n$th energy interval. It is equalized to $nT$, because it corresponds to the $n$th stable parameter range of $\lambda^{\prime}$. Then, $|E_n-E_c|=\exp(-nT/\alpha)\sim (T/\alpha)^{-n}\sim (1/2.37)^{n}$.  This exponent 2.37 is consistent with the numerically obtained value 2.4 in Fig.~4(c). 
Note that the exponent is determined only by $\alpha$ and $\omega_2$, which are further determined only by $\partial^2\Delta U/\partial x^2$ and $\partial^2\Delta U/\partial y^2$ at the saddle point. Thus, we find that the simple approximation by Eq.~(5) explains the singular behavior rather well.  

\section{Simultaneous jump motions of a small number of particles}
\subsection{Four particles in a square box} 
The anomalous behavior in the two-particle system near the delocalization transition is due to the coexistence of the KAM tori and chaos.
If the degree of the freedom of motion is increased, the regions of the KAM tori are expected to become smaller, and therefore, the effect of the anomalous behavior is also expected to become smaller. However, there is another peculiar behavior in chaotic diffusion by a number of particles near the delocalization transition.
We have performed numerical simulations of four particles in a square box of $L=1.4$ under the periodic boundary conditions. No special symmetry is further assumed. Figure 6(a) is a trajectory $(x_1,y_1)$ of the first particle at $E=0.89$. The chaotic diffusion occurs in the $(1,1)$ or $(1,-1)$ direction. The unit cell is a square of size $L/8$ in this system.
Figure 6(b) displays time evolutions of $x_i(t)$  for $i=1,2,3$ and 4 for $E=0.89$. Jump motions toward the neighboring cells are clearly seen.  
The jump motion occurs almost simultaneously for the four particles. The first particle comes into the unit cell of the second particle, then, the second particle needs to go away from the original unit cell and moves into the unit cell of the third particle.  Successively, the third particle comes into the unit cell of the fourth cell particles, and the fourth particle invades into the unit cell of the first particle. We call this type of successive jump motions a simultaneous jump motion in this paper.  It reminds of the heterogeneous and intermittent motion of particles near the glass transition or the jamming transition.  The jump motion occurs as a swirling motion, because the energy $E$ is close to the critical energy $E_c$. 

\begin{figure}[tbp]
\includegraphics[height=4.cm]{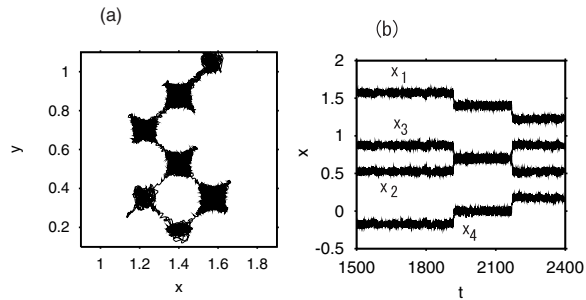}
\caption{(a) Trajectory of the first particle at $E=0.89$ in a four-particle system in a square box of $L=1.4$. (b) Time evolutions of $x_i(t)$ for $i=1,2,3$ and 4 at $E=0.89$.}
\label{fig6}
\end{figure}
\begin{figure}[tbp]
\includegraphics[height=4.cm]{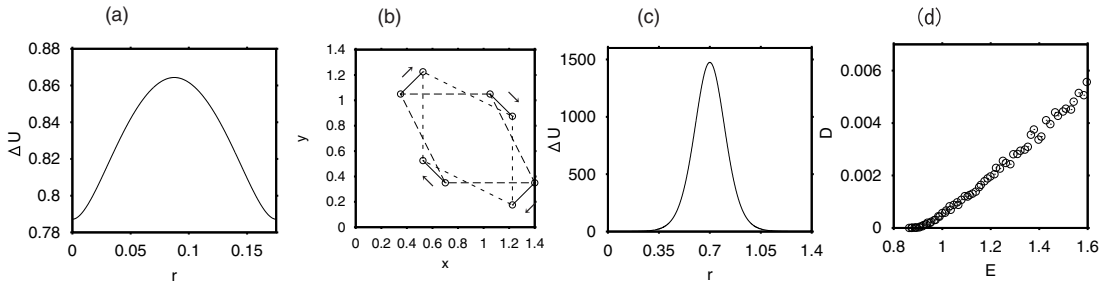}
\caption{(a) Potential energy $\Delta U$ as a function of the simultaneous displacement by $r$ of the four particles along the $(1,1)$ or $(1,-1)$ direction. (b) Displacement of the four particles for the potential $\Delta U$ in Fig.~7(a). Four particles move in the directions of the arrows. (c) Potential energy $\Delta U$ as  a function of the single-particle displacement by $r$ along the $(0,-1)$ direction. (d) Numerically evaluated diffusion constant as a function of $E$.}
\label{fig7}
\end{figure}
Figure 7(a) displays the potential energy $\Delta U$, when the four particles move simultaneously as $(x_1,y_1)=(3L/4+r,3L/4-r), (x_2,y_2)=(2L/4-r,L/4+r), (x_3,y_3)=(2L/4+r,3L/4+r)$ and $(x_4,y_4)=(L-r,L/4-r)$. 
Here, the potential energy $\Delta U$ is calculated as the sum of the potential energy of  the four particles measured from $U(r_0)$.  When $r=L/16$, the potential $\Delta U$ takes the maximum. The point is a saddle point because it is a minimum point for the potential energy along the transverse direction. When $r=L/8$, the potential energy takes the same value as the initial configuration at $r=0$, because the initial and the final configurations of the four particles are the same parallelogram, although their directions are different as shown in Fig.~7(b). Each particle moves by one unit cell by making the course shown in Fig.~7(b). The configuration at $r=L/16$ is considered to be a saddle point. 
The minimum energy to go over the saddle point is $E=0.8643$. In direct numerical simulation, the chaotic diffusion appeared for $E>0.875$. The critical energy for the delocalization transition is close to the energy of the saddle point. If the total energy $E$ is close to the energy of the saddle point, it is necessary for the four particles to exhibit the successive motion as shown in Fig.~7(b) to go over the saddle point. 
It takes much larger energy for a single particle to go over the barrier, setting the other particles fixed in the original position.  Figure 7(c) displays the change of the potential energy $\Delta U$ for such a course $(x_1,y_1)=(3L/4,3L/4-r), (x_2,y_2)=(2L/4,L/4), (x_3,y_3)=(2L/4,3L/4)$ and $(x_4,y_4)=(L,L/4)$. When only the first particle is moved downward by $L=1.4$, an original configuration is recovered owing to the periodic boundary conditions. However, the critical energy is about 1475. Such a jump motion cannot occur if the energy $E$ is around 1. Figure 7(d) displays the diffusion constant $D$ for one particle
calculated from $\langle (x(t)-x(0))^2+(y(t)-y(0))^2\rangle/(4t)$.  The diffusion constant increases roughly in proportion to $E-E_c$. 
\begin{figure}[tbp]
\includegraphics[height=5.5cm]{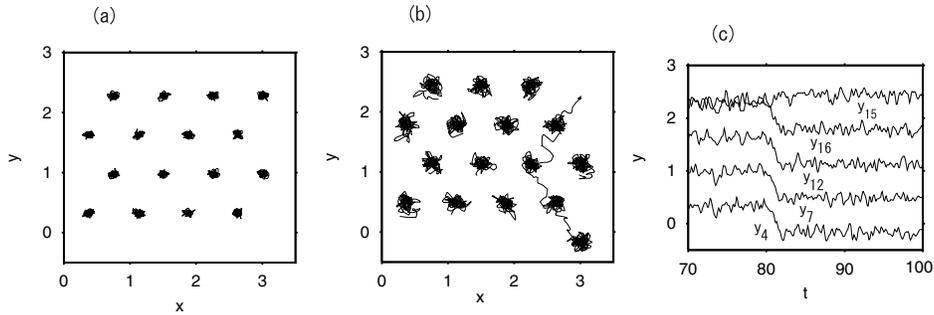}
\caption{(a) Trajectories of sixteen particles at $E=3.827$ between $800<t<1000$. (b) Trajectories of sixteen particles at $E=8.158$ between $800<t<1000$. (c) Time evolution of $y_i(t)$ for $i=4,7,12,15$ and 16 at $E=8.158$.}
\label{fig8}
\end{figure}
\begin{figure}[tbp]
\includegraphics[height=5.5cm]{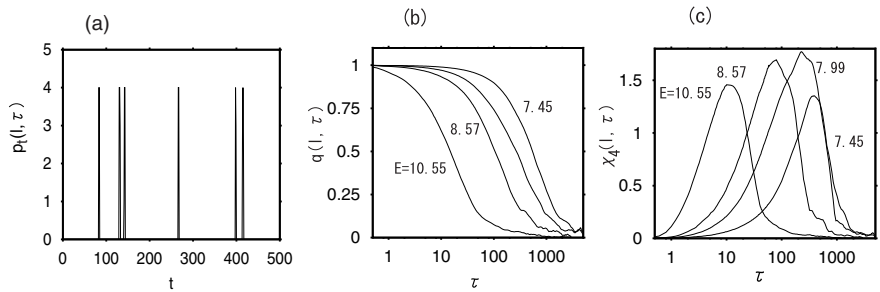}
\caption{(a) Time evolution of $p_t(l,\tau)$ for $l=0.55$ and $\tau=4$ at $E=8.158$. (b) $q(l,\tau)$ as a function of $\tau$ for $l=L/8=0.375$  at several $E$'s. (c) $\chi_4(l,\tau)$ as a function of $\tau$ for $l=0.375$ at several $E$'s.
}
\label{fig9}
\end{figure}
\begin{figure}[tbp]
\includegraphics[height=5.cm]{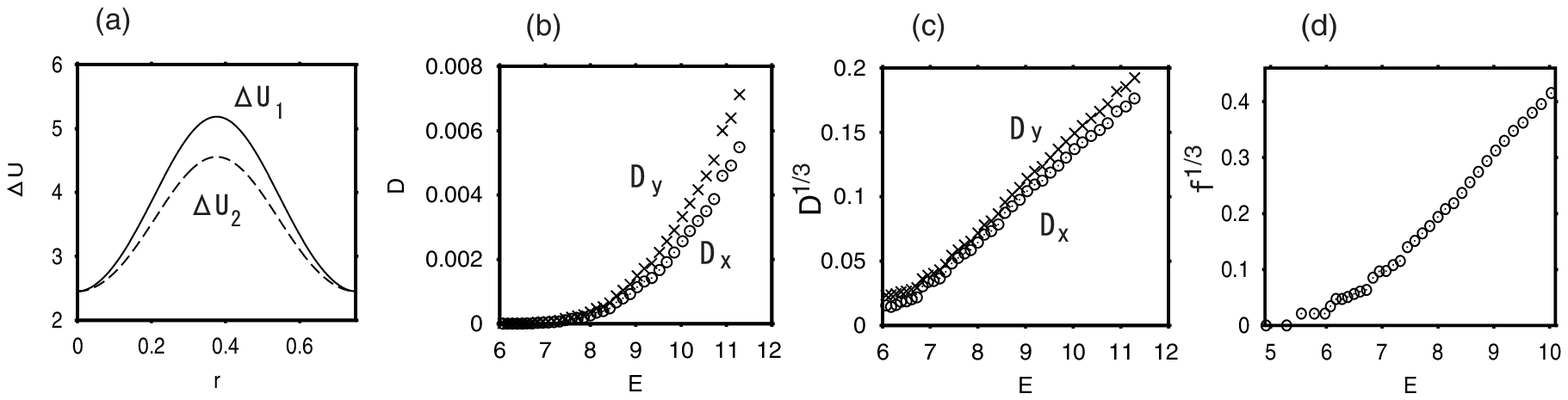}
\caption{(a) Potential energy $\Delta U_1$ (solid curve) along a zigzag displacement of four particles, $\Delta U_2$ along a horizontal displacement. (b) Diffusion constants $D_x$ and $D_y$ as a function of $E$. (c) $D_x^{1/3}$ and $D_y^{1/3}$ as a function of $E$. (d) $f^{1/3}$ as a function of $E$, where $f$ is the frequency of the jump motion.}
\label{fig10}
\end{figure}
\subsection{Sixteen particles in a rectangular box}
Next, we consider sixteen particles in a rectangular box of size $L_x\times L_y=L\times (\sqrt{3}/2)L$ under the periodic boundary conditions. The size $L$ is fixed to be 3.008. A triangular lattice is a natural configuration for the solid phase in this system. Figure 8(a) displays trajectories of the sixteen particles, which form a typical triangular lattice for $E=3.827$. Figure 8(b) displays trajectories of the sixteen particles at $E=8.158$ between $800<t<1000$. String-like trajectories appear near $x=2.5$, which imply successive jump motions.
Figure 8(c) displays time evolutions of $y_i(t)$ for $i=4,7,12,15$ and 16. Particles $i=4,7,12,16$ jump downward by one layer almost simultaneously, and the other particles are confined in the original cells in this time interval. 
The jump motion occurs like a chain-reaction in the $y$-direction owing to the periodic boundary conditions. This is another example of the simultaneous jump motions. In this sixteen-particle system, the simultaneous jump motion by all the sixteen particles rarely occurs near the critical energy in contrast to the four-particle system. 

We try to characterize the jump motions using some quantities such as $q(l,\tau)$ and $\chi_4(l,\tau)$, which were used to characterize the heterogeneous dynamics near the glass transition or the jamming transition~\cite{rf:15,rf:16,rf:17}. 
Figure 9(a) displays time evolution of $p_t(l,\tau)=\sum \theta[\{(x_i(t)-x_i(t-\tau))^2+(y_i(t)-y_i(t-\tau))^2\}^{1/2}-l]$ for $l=0.55$ and $\tau=4$ at $E=8.158$, where $\theta(x)$ is the Heaviside step function. The quantity $p_t(l,\tau)$ represents the number of particles whose displacement in the time interval $\tau$ is larger than $l$. Figure 9(a) implies that the simultaneous jump motion by four particles occurs intermittently. The four particles are not fixed but are chosen randomly.   Simultaneous jump motions by eight particles or sixteen particles do not appear in the time sequence. Figure 9(b) displays $q(l,\tau)=\langle q_t(l,\tau)\rangle$ as a function of $\tau$ for several values of $E$ for $l=L/8$, where $q_t(l,\tau)=(1/16)\sum \theta[l-\{(x_i(t)-x_i(t-\tau))^2+(y_i(t)-y_i(t-\tau))^2\}^{1/2}]$ and $\langle\cdots\rangle$ implies the time average. The quantity $q_t(l,\tau)$ represents the number ratio of particles whose displacement in the time interval $\tau$ is smaller than $l$.  The quantity $q(l,\tau)$ is the temporal average of $q_t(l,\tau)$.
The time scale, where $q(l,\tau)$ decreases rapidly from 1 to 0, increases as $E$ approaches $E_c$. The quantity $q(l,\tau)$ characterizes that the chaotic jump motions become more intermittent as $E\rightarrow E_c$.     
Figure 9(c) displays $\chi_4(l,\tau)$ as a function of $\tau$ for several values of $E$  for $l=L/8$, where $\chi_4(l,\tau)$ is defined as 
\[\chi_4(l,\tau)=(1/16)[\langle q_t(l,\tau)^2\rangle-\langle q_t(l,\tau)\rangle^2].\]
 The quantity $\chi_4(l,\tau)$ take a maximum value at a certain time, and the time scale increases as $E$ is decreased to $E_c$. This quantity also characterizes the intermittency of the simultaneous jump motions.   

Figure 10(a) displays the potential energy $\Delta U_1$ (solid curve) along the  zigzag displacement of four particles:
$(x_4,y_4)=(7L/8+r/2,\sqrt{3}/2(L/8-r)), (x_7,y_7)=(6L/8+r/2,\sqrt{3}/2(3L/8-r)),(x_{12},y_{12})=(7L/8-r/2,\sqrt{3}/2(5L/8-r)),(x_{16},y_{16})=(L-r/2.\sqrt{3}/2(7L/8-r))$, and the potential energy $\Delta U_2$ (dashed curve) along the horizontal displacement $(x_1,y_1)=(L/8-r,(\sqrt{3}/2)L/8),(x_2,y_2)=(3L/8-r,(\sqrt{3}/2)L/8),(x_3,y_3)=(5L/8-r,(\sqrt{3}/2)L/8),(x_4,y_4)=(7L/8-r,(\sqrt{3}/2)L/8)$. The other particles are fixed to the original positions. 
The zigzag displacement corresponds the simultaneous jump motion of four particles shown in Fig.~8(b). 
The potential energy at the peak position is smaller for $\Delta U_2$ than for $\Delta U_1$.  It implies that the horizontal motion is easier to occur. However, the zigzag displacement appears more often in direct numerical simulation as shown in Fig.~8(b).  We do not understand the reason well yet, however, it is partly because the configuration number of the zigzag displacement is larger than the horizontal displacement. That is, the configuration number of the horizontal displacements is 4 because one of the four layers moves, however, the combination number of the zigzag displacement is counted as $4\times 2\times 2=16$. 

Figure 10(b) displays two diffusion constants, i.e., $D_x$ in the $x$-direction and $D_y$ in the $y$-direction, as a function of $E$. 
The diffusion constants $D_x$ and $D_y$ are slightly different in this numerical simulation  because of the anisotropy in this system, however, their behavior  is very similar near the critical energy.   
Figure 10(c) displays $D_x^{1/3}$ and $D_y^{1/3}$ as a function of $E$. The diffusion constants increase roughly as $D_{x,y}\sim (E-E_0)^{3}$  for $E>6.8$ where $E_0\sim 6$. When $E<6.8$, the jump motion becomes very rare, however the jump frequency is not zero for $E>E_c\sim 5.2$. Here, $E_c$ is estimated by numerical simulation performed until $t=100000$, and it might be lowered in a even longer simulation. The critical energy for the delocalization transition is comparable to the maximum energy of $\Delta U$ shown in Fig.~10(a). 
Figure 10(d) displays $f(E)^{1/3}$, where $f(E)$ is the frequency of the jump motion per unit time, as a function of $E$. The frequency $f$ of the jump motion also increases as $f\sim (E-E_0)^{3}$ for $E>6.8$, although the reason of the exponent $3$ is not understood yet.

\section{Summary and discussion}
We have performed numerical simulations of Hamiltonian systems with repulsive interaction, and studied the delocalization transition. 
We have found that the dynamics around the saddle point is very important near the delocalization transition, because the saddle point plays a role of the lowest barrier between the neighboring cells. In a two-particle system, we have found infinite alternation of the stability and the instability of the straight motion passing through the saddle point. It is closely related to the anomalous behavior of the jump frequency $f$ above the critical energy. In a four-particle system in a square box and a sixteen-particle system in a rectangular box, we have found that simultaneous jump motions occur intermittently near the transition point. If each particle is located deeply inside of each cell, the chaotic motion for the particle is weakly correlated from that for the other particles. However, it is necessary for several particles to exhibit jump motion almost simultaneously to go over the saddle point using a small amount of energy.  All the four particles exhibit the simultaneous jump motion in the four-particle system. On the other hand, four particles exhibit the simultaneous jump motion near the delocalization transition in the sixteen-particle system.  These simulatenous jump motions might be related with the heterogeneous dynamics near the glass transition and the jamming transition, although the detailed relation with such systems of a large number of particles is not known well. 

There is a definite critical energy $E_c$ of the delocalization transition in Hamiltonian systems of a small number of particles, below which each particle is completely confined in each unit cell. The critical energy seems to be determined by the energy of the lowest saddle point. We consider that the critical energy of the delocalization transition in the Hamiltonian systems of a small number of particles is generally different from the critical energy for the statistical-mechanical liquid-solid phase transition in an infinitely large system. It is partly because a long-range order needs to appear against the statistical-mechanical fluctuations below the liquid-solid phase transition point. For example, the long-range order does not appear in two dimensions owing to the long-wavelength fluctuations~\cite{rf:18}.  The long-wavelength fluctuations are not directly related to the energy of the saddle points. However, the detailed relation of the delocalization transition and the statistical-mechanical liquid-solid phase transition is not understood yet but it is left as a future problem.

\end{document}